\documentclass[12pt,preprint]{aastex}

\shorttitle{Hard X-ray and UV Two-ribbon Flare}
\shortauthors{Cheng et al.}

\begin{document}

\title{Hard X-ray and UV Observations of the 2005 January 15 Two-ribbon Flare}
\author{J. X. Cheng\altaffilmark{1,2}, G. Kerr\altaffilmark{3}, J. Qiu\altaffilmark{1} }
\affil{$^1$ Department of Physics, Montana State University,
Bozeman, MT 59717-3840, USA} \affil{$2$ School of Astronomy $\&$
Space Science, Nanjing University, Nanjing 210093, China}
\affil{$^3$ SUPA, School of Physics and Astronomy, University of
Glasgow, Scotland, U.K} \email{jxcheng@nju.edu.cn}

\begin{abstract}
It is well known that two-ribbon flares observed in H$\alpha$ and
ultraviolet (UV) wavelengths mostly exhibit compact and localized
hard X-ray (HXR) sources \citep{warren01}. In this paper, we
present comprehensive analysis of a two-ribbon flare observed in
UV 1600\AA\ by TRACE and in HXRs by RHESSI. HXR (25 - 100 keV)
imaging observations show two kernels of size (FWHM) 15\arcsec
moving along the two UV ribbons. We find the following results.
(1) UV brightening is substantially enhanced wherever and whenever
the compact HXR kernel is passing, and during the hard X-ray
transit across a certain region, the UV counts light curve in that
region is temporally correlated with the hard X-ray total flux
light curve. After the passage of the HXR kernel, the UV light
curve exhibits smooth monotonical decay. (2) We measure the
apparent motion speed of the HXR sources and UV ribbon fronts, and
decompose the motion into parallel and perpendicular motions with
respect to the magnetic polarity inversion line (PIL). It is found
that HXR kernels and UV fronts exhibit similar apparent motion
patterns and speeds. The parallel motion dominates during the rise
of the HXR emission, and the perpendicular motion starts and
dominates at the HXR peak, the apparent motion speed being 10 - 40
km s$^{-1}$. (3) We also find that UV emission is characterized by
a rapid rise correlated with HXRs, followed by a long decay on
timescales of 15 - 30 min. The above analysis provides evidence
that UV brightening is primarily caused by beam heating, which
also produces thick-target HXR emission. The thermal origin of UV
emission cannot be excluded, but would produce weaker heating by
one order of magnitude. The extended UV ribbons in this event are
most likely a result of sequential reconnection along the PIL,
which produces individual flux tubes (post-flare loops),
subsequent non-thermal energy release and heating in these flux
tubes, and then the very long cooling time of the transition
region at the feet of these flux tubes.
\end{abstract}


\section{INTRODUCTION}
Solar flares are impulsive energy release events in the solar atmosphere.
During flares, magnetic free energy is released by magnetic reconnection to
accelerate particles, heat plasmas, and drive mass motions.
Accelerated particles interacting with the lower solar atmosphere produce
electromagnetic radiation at optical, ultraviolet (UV), and hard X-ray (HXR)
wavelengths. By using multi-wavelength observations obtained from space telescopes, we can
study various physical processes in flares.

Many studies have been focused on the temporal and spatial relationship between
HXR and UV continuum emissions. \citet{kane71} used HXR data from OGO satellites
and ground-based sudden frequency deviation at 10-1030 \AA, and showed similar
time profiles at both wavelengths, particularly a strong temporal relationship
during the rise of the flare. These studies provided observational evidence that HXR
and UV emissions are associated with a common origin. Using data
from SMM, \citet{cheng88} further studied the timing of HXR and UV broadband emissions, finding a strong
temporal correlation between the two. They suggested that UV and HXR emissions are most likely produced by the
same source particle population. They also provided evidence of localized UV and HXR sources through the UVSP images.
\citet{warren01} suggested that the UV emission tends to exist in more extended ribbons while the HXR emission is generally more localized.
\citet{alexander06} studied the flare on 2002 July 16 and concluded that the UV and HXR emissions
are directly associated with the same flare energy release process, although the spatial separation existed.
Furthermore, \citet{coyner09} conducted a statistical study focused on the relationship
between the localized sources at both wavelengths and indicated two distinct types of UV emissions,
one correlated with the HXR emission and the other more likely associated with a thermal origin.

The missing hard X-ray two-ribbon is a myth in solar flares. There
are several possible reasons for the lack of HXR ribbons. First,
it is suggested that energy release along the UV and optical
ribbon is not uniform, and the dynamic range of the present HXR
image reconstruction is about 10 \citep{sui04}, so energy
deposition rate below one tenth of the maximum cannot be recovered
in reconstructed HXR images. Second, either due to non-uniform
energy release rate or due to the manner of energy release,
namely, heating versus particle acceleration, particle
acceleration may prefer to be localized. And last, energy release
itself is localized and the UV or optical two-ribbons are a
manifest of localized energy release sequentially along the ribbon
and the elongated ribbon cooling time in UV or H$\alpha$
wavelength. In the last scenario, both of the hard X-ray and UV
continuum emissions are thought to result from direct particle
injection in the chromosphere. {\bf It has come to the recognition
that even a two-ribbon flare is not 2-dimensional, but consists of
numerous flare loops formed and energized individually by magnetic
reconnection. Distinguishing the three different scenarios will
provide important information of the form of energy release,
namely, by direct heating or by particle acceleration, in these
individual flare loops during the flare. Ideally, the best
observational approach to distinguish the three scenarios is to
spatially resolve electron deposit along flare loops, which may
only have a cross-sectional area of around 1\arcsec. However,
existing hard X-ray imaging capabilities are not able to resolve
with such accuracy. Alternatively, high-resolution optical and UV
imaging observations in combination with hard X-ray observations
may be used to provide some spatial information of hard X-ray
emissions.} In this paper, we conduct a comprehensive quantitative
analysis in order to shed light on this issue. The remainder of
this paper is organized as follows. In \S 2 we briefly discuss the
data analysis. We present the  results in \S3. Discussions and
conclusions are made in \S 4.

\section{Observations and Data Analysis}
We present analysis of HXR and UV observations of an X2.6 flare on 2005 January 15.
 According to $Solar$ $Geophysical$ $Data$, the flare occurred from 22:25 UT to 23:31 UT
in NOAA 720 when the active region was at the disk center. The
longitudinal magnetograms of the active region were obtained by
$Michelson$ $Doppler$ $Imager$ \citep[MDI;][]{scherrer95}. The
flare was observed, shortly after its onset, by the $Transition$
$Region$ $and$ $Coronal$ $Explorer$ \citep[TRACE;][]{handy99}
 in 1600 \AA\ ultraviolet (UV) continuum with the best cadence (2~s) of the instrument and a pixel scale of 0.5\arcsec.
Observations at this wavelength reflect the flare emission in the lower atmosphere, or emission at the feet of flaring loops.
The $Reuven$ $Ramaty$ $High$ $Energy$ $Solar$ $Spectroscopic$ $Imager$
\citep[RHESSI][]{lin02} also observed the flare at X-ray wavelengths for its entire duration.

From RHESSI observations, we reconstruct HXR images with the PIXON
method \citep{metcalf96} in two energy ranges of 25-50 keV and
50-100 keV, to capture the non-thermal emission of the flare.
Altogether 322 HXR images are reconstructed from 22:30:46.894 UT
to 23:13:26.894 UT. The field of view of the reconstructed HXR
maps is 256\arcsec $\times$ 256\arcsec with the spatial resolution
2\arcsec $\times$ 2\arcsec. The integral time to make HXR maps is
taken to be 4 s, 8 s, or 12 s, depending on the observed counts
rate during which the total integrated count rate is greater than
3000. The RHESSI maps and TRACE images are both coaligned with an
SoHO/MDI magnetogram obtained before the flare.

\begin{figure}
\epsscale{1.0} \plotone{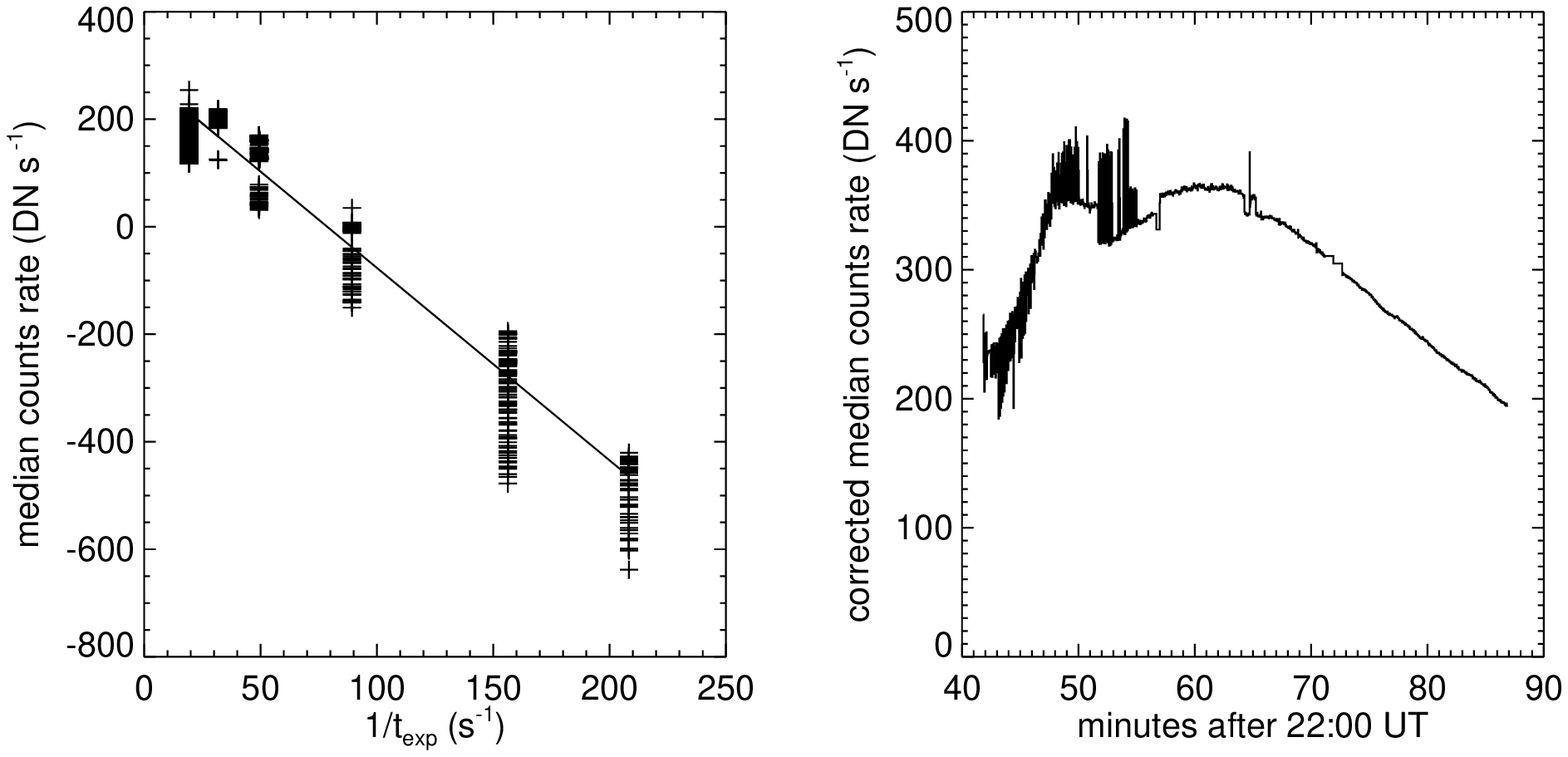} \caption{Left: the plot of
uncorrected median values of UV counts rate versus 1/$\tau_{exp}$
(see text). Right: corrected UV median counts rate light
curve.}\label{median}
\end{figure}

To derive semi-quantitative information of UV emission, we look
into the calibration of TRACE images. The TRACE 1600\AA\ images
are first processed using the ${\rm trace\_prep.pro}$ built in the
SolarSoftWare (SSW), which performs dark current and flat field
correction and exposure normalization. Figure~\ref{median},
however, shows that the median $I_m(t)$ of the processed UV
images, which is dominated by non-flaring quiescent regions, takes
negative values at short exposures. Furthermore, the figure shows
that $I_m(t)$ is inversely proportional to the exposure time
$\tau_{exp}$, indicating that a dark current pedestal is not
properly removed from the initial processing. We then obtain the
dark current pedestal as the slope of the linear fit to the
$I_m(t) \sim 1/\tau_{exp}$ scatter plot and remove it from the UV
images \citep{Qiu10}. With this first-order correction, the median
of the UV emission is recovered, which varies between 250 and 350
counts per second. It is seen that there are still fluctuations in
the corrected median light curve. We then discard the frames of UV
images whose corrected median is different from the mean median by
more than 2$\sigma$, where $\sigma$ is the standard deviation of
the $I_m(t) \sim 1/\tau_{exp}$ fit. Such selection leaves 1313
 out of 1350 frames of images for further analysis, so
the cadence of the UV observations is not compromised. We further
normalize each UV image to the median, namely, the UV counts rate
in TRACE UV images is measured as how many times the quiescent
median. {\bf Note that this first-order correction of the dark pedestal
offset is not perfect, so there are still some remaining artifacts
in the corrected light curves, such as some very short-lived spikes
and dips in some low-count regions during the flare maximum.}

Figure~\ref{overview} gives snapshots of the flare observed in UV and hard X-rays. TRACE 1600\AA\ images show
two flare ribbons along the magnetic polarity inversion line. The ribbon in the negative magnetic
field (N-ribbon) appears to consist of two sections. The section in the west is brightened
first, and the section in the east emits more strongly after 23:00 UT. HXR emission at 50-100
keV is primarily produced in one compact source S1, or hard X-ray kernel, located in the negative magnetic field.
A very weak source S2, whose intensity is only about 10\% of that of the strong source S1, is also visible in the positive
magnetic field. The kernels are seen to ``move" along the UV ribbon during the flare and are located at
where UV emission is strong. Hard X-ray maps at 25-50 keV (not shown in the paper) show the same morphology
and evolution pattern as the 50-100 keV sources. Therefore, in this flare, hard X-ray emission above 25 keV is considered
to be thick-target non-thermal emission from the foot-points of flaring loops.

The bottom panel of the figure shows the corrected UV total counts light curve together with the RHESSI hard X-ray light
curves at 50 - 100 keV, and the GOES soft X-ray light curve at 1-8 \AA.
It is seen that UV and HXRs rise, peak, and decay nearly simultaneously, suggesting the common
origin of emissions at the two wavelengths. With the high cadence observations at the two wavelengths,
we conduct detailed analysis to examine the temporal and spatial relationship between hard X-ray and UV
emissions and the evolution pattern of hard X-ray and UV emissions.

\section{Results}
\subsection{UV and HXR emissions}
\begin{figure}
\epsscale{0.90} \plotone{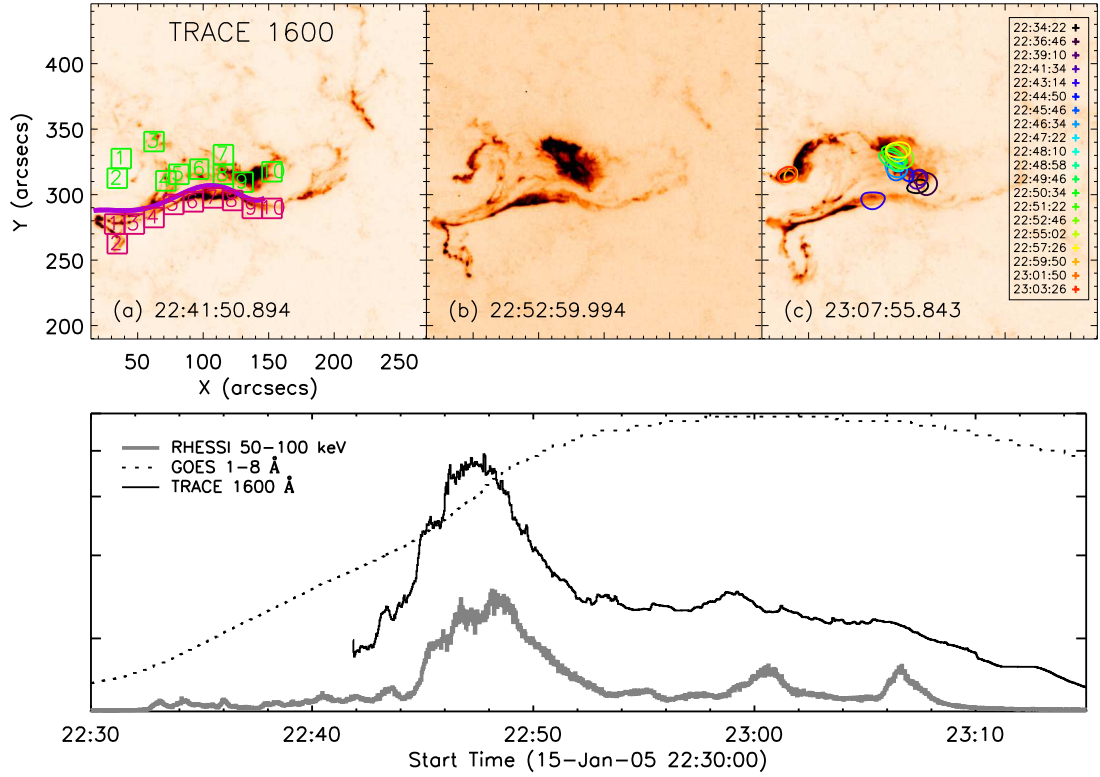} \caption{Upper: Snapshots of the
X2.6 flare observed by TRACE 1600 \AA\ on 2005 January 15. The
magnetic polarity inversion line is outlined by the purple line in
the left panel. The boxes in the left panel indicate regions where
spatially resolved UV counts rate light curves are obtained. In
the right panel, the RHESSI HXR maps in 50-100 keV are overlayed.
Bottom: Light curves for the flare in different energy
bands.}\label{overview}
\end{figure}

\begin{figure}
\epsscale{.80} \plotone{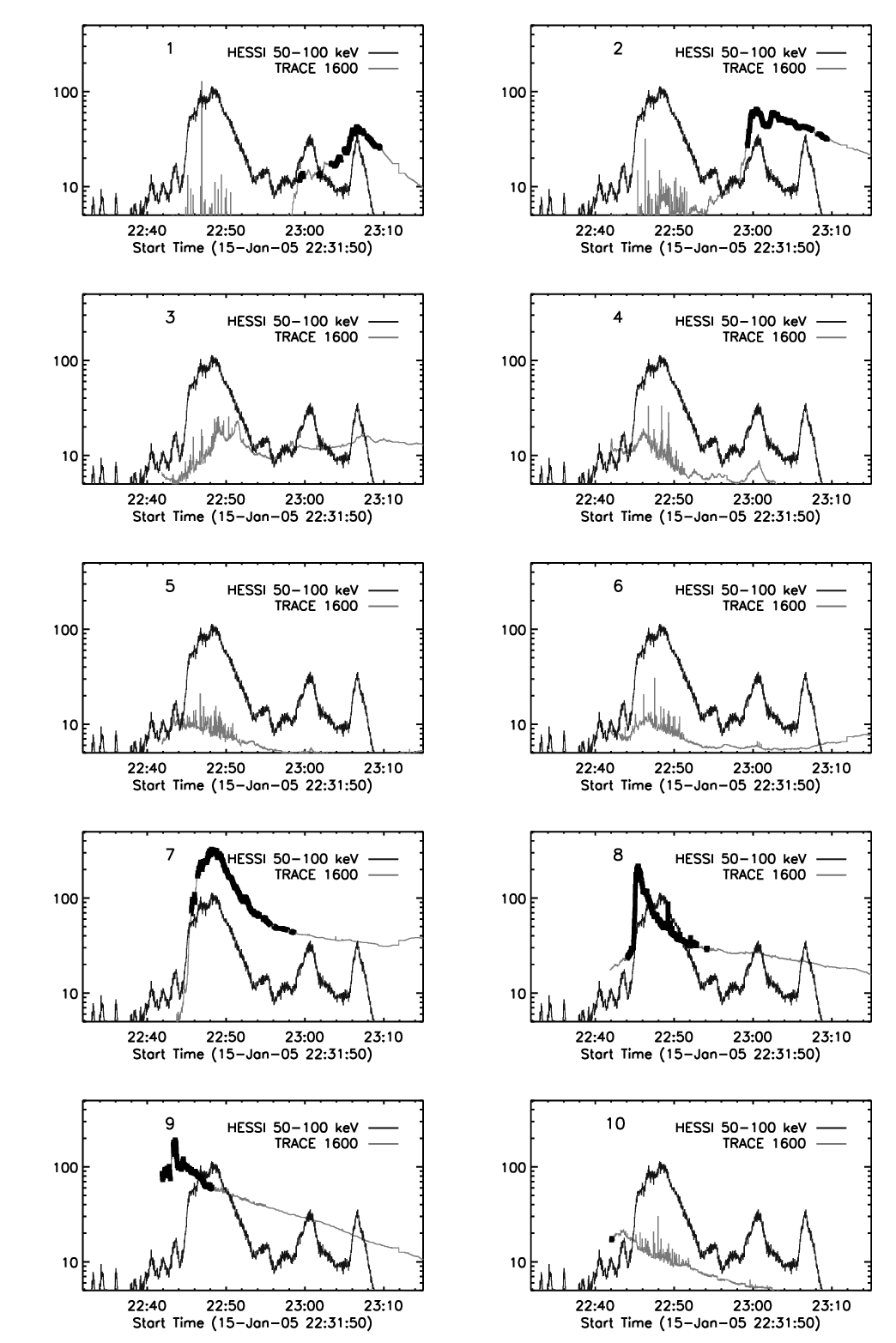} \caption{The corrected TRACE UV
counts rate light curves (grey) in small boxes as indicated in
Figure 2 (N-ribbon), compared with HXR 50-100 keV count flux light
curve (dark) arbitrarily normalized. The thick dark color on a
TRACE light curve shows the timing when the HXR source passes
through a certain box.}\label{box}
\end{figure}

Figure~\ref{overview} shows two flare ribbons in UV images and two
kernels in hard X-ray emission. In order to investigate the
spatial relationship between UV and HXR emissions, we compare the
hard X-ray light curve with UV light curves in spatially resolved
regions along the UV ribbons. For this purpose, we divide the UV
flare ribbons into small boxes, each of size 15\arcsec $\times$
15\arcsec. This is comparable with the FWHM of the HXR kernels
which are approximately circular-shaped. These boxes are displayed
in Figure~\ref{overview}. The boxes are selected in a left-right,
top-bottom sequence, and different colors indicate boxes along the
positive and negative ribbons, respectively. We compute the mean
counts rate, normalized to the quiescent median, in each small box
to obtain the TRACE light curves. The comparison of HXR and
spatially resolved UV light curves is shown in Figure~\ref{box}.
In the figure, we denote on the UV light curve by thick dark lines
the duration when the centroid of the HXR kernel falls into the
box.

Figure~\ref{box} shows that, along both ribbons, when the HXR kernel passes through a specific part (box)
of the UV ribbon, the UV light curve in that box rises rapidly with significant enhancement by one to two orders of magnitude
over the pre-flare quiescent background. During the passage of the HXR kernel, the UV light curve in
the box is well correlated with the HXR light curve. The peak UV
enhancement in individual boxes is in general scaled with the HXR flux
(except for box 9). If the HXR flux is very strong, there is a strong UV enhancement,
otherwise, we observe a weak enhancement in the UV light curve.
In each box, after the passage of the HXR kernel, the UV light curve exhibits a smooth monotonical decay.

In the negative magnetic field, at the beginning, the HXR kernel
S1 is located at the western section of the N-ribbon. After 23:00
UT, the HXR source S1 shifts to the eastern section of the
N-ribbon with much weaker emission. The HXR emission exhibits two
small bursts after 23:00 UT. One is around 23:00 UT, another is at
23:07 UT, both coincident with brightened UV regions. We observe a
similar pattern in the positive ribbon. When the HXR kernel S2
passes through a certain UV region, the UV light curve in that box
shows significant enhancement, and it decays smoothly after the
passage of S2. Note that the magnitude of UV enhancement in the
P-ribbon is much smaller than in the N-ribbon by an order of
magnitude, consistent with the relative intensities of two hard
X-ray kernels.

The above comparison of hard X-ray light curve with spatially
resolved UV light curves suggest that, along the ribbons, wherever
and whenever HXR kernel passes, significant brightening will occur
in the UV emission. Therefore, enhanced UV emission is primarily
produced by precipitating electrons that also produce thick-target
hard X-rays.

\subsection{Evolution of HXR kernels and UV ribbons}
The previous section shows that hard X-ray kernels move along the
UV ribbons, and UV emission is strongest at the location of hard
X-ray kernels. Such apparent motion is caused by reconnection and
subsequent energy release along adjacent field lines along the
ribbons. In this section, we examine in detail the apparent motion
pattern of the energy deposit sites using hard X-ray as well as UV
imaging observations.

Following the approach by \citet{qiu09},  we quantitatively
characterize evolution of the flare ribbons and kernels with
respect to the PIL. For this purpose, we first determine the
profile of the PIL from the magnetogram, which is curved and
extended nearly in the east to west direction. We then decompose
the spread of ribbon brightening into two directions, parallel
(elongation) and perpendicular (expansion) to the local PIL. To
quantify the elongation and expansion of flare ribbons, we measure
the following quantities for each ribbon at each time frame: the
entire ribbon length ($l_{\|}$) projected along the PIL, and the
mean distance ($d_\bot$) of the ribbon front perpendicular to the
local PIL. The mean perpendicular distance $d_\bot$ is computed as
$d_\bot$ = $S/l_\|$, where $S$ is the total area enclosed between
the outer edge of the ribbon and the section of the PIL along the
ribbon. The time profile of $l_{\|}$ gives a general description
of the ribbon length growth along the PIL, whereas the time
evolution of $d_\bot$ shows the separation of ribbon front away
from the PIL.

In the similar way, we also decompose the trajectory of the
centroid of each HXR kernel as components parallel and
perpendicular to the PIL, respectively. The parallel distance of
the centroid is measured from the same reference point where the
UV ribbon starts, while the perpendicular distance refers to the
distance of the centroid away from the PIL.

\begin{figure}
\epsscale{.90} \plotone{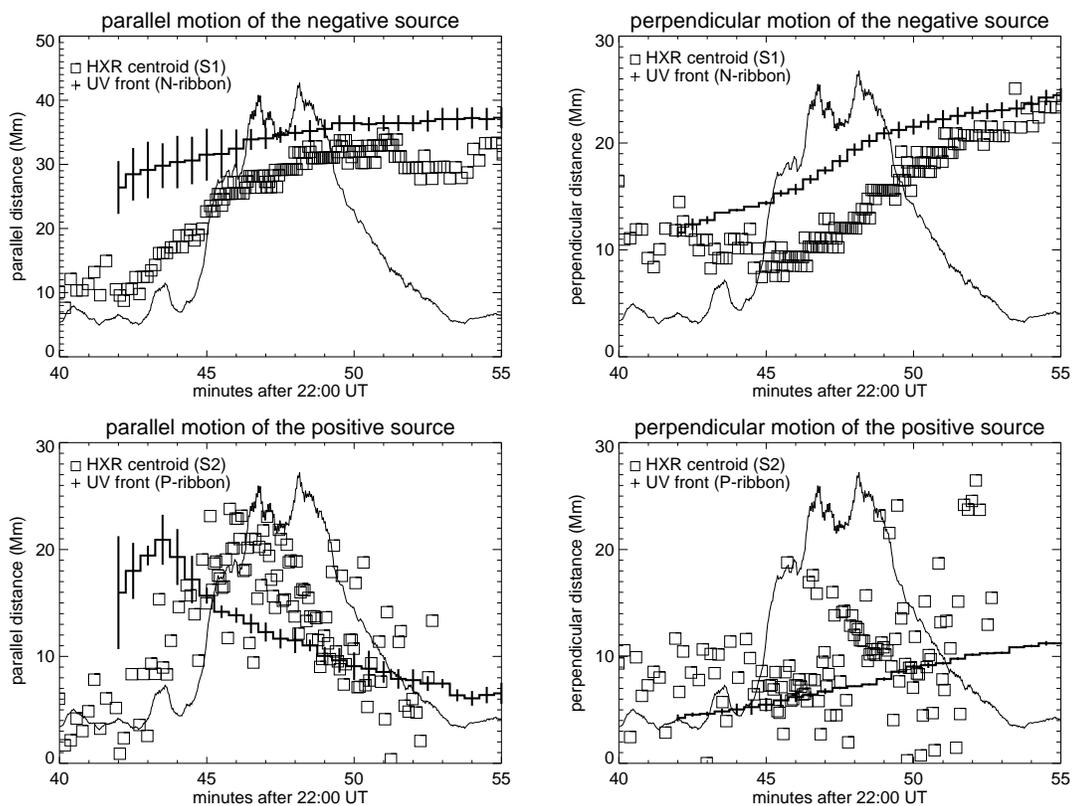} \caption{Parallel (left panels)
and perpendicular (right panels) distances of the HXR kernels and
UV ribbon fronts with respect to the polarity inversion line. The
upper panels show the motion of the HXR kernel S1 and UV N-ribbon,
and the lower panels show the motion of the HXR kernel S2 and UV
P-ribbon. The solid dark line shows the hard X-ray light curve at
100 - 300 keV as a reference.}\label{motion}
\end{figure}

Figure~\ref{motion} shows the parallel and perpendicular motion of the UV ribbon fronts as well as the HXR kernels.
 The analysis of the apparent ribbon motion is more
ambiguous for the western section of the N-ribbon, which is
brightened significantly after 23:00 UT, hence we only present the
result for the eastern section of the N-ribbon before 23:00 UT.
The N-ribbon elongates eastward along the PIL and moves away from
the PIL, which is very well correlated with the motion of the HXR
kernel S1. The maximum length of the negative UV ribbon ($l_{\|}$)
can reach 40 Mm, with the average elongation speed of 40 km
s$^{-1}$ from 22:42 UT to 22:49 UT. After that, the ribbon
elongation slows down. The mean perpendicular distance (d$_\bot$)
of the ribbon front varies from 14 Mm to 27 Mm with a mean speed
of 22 km s$^{-1}$ between 22:43 UT to 22:49 UT and 8 km s$^{-1}$
afterwards.

The HXR kernel motion exhibits very similar trend. From 22:43:00 UT to 22:50:00 UT, the source S1 moves along the
PIL from 7 Mm to 25 Mm with an average speed about 45 km s$^{-1}$.
After that, there is no systematic parallel motion any more.
The perpendicular motion starts several minutes later than the parallel motion.
It begins at about 22:46 UT and lasts until  23:00 UT.
The perpendicular motion can be divided into two phases.
Before 22:52 UT, the perpendicular distance varies from 7 Mm to 15 Mm with a average speed about 22 km s$^{-1}$. After that,
the perpendicular motion becomes slower and in the following  7 minutes, it moves only 5 Mm with a mean speed  about 12 km s$^{-1}$.

The positive UV ribbon locates at the south of the PIL. Its
trajectory is different from that of the negative ribbon. At the
beginning, it spreads eastward along the PIL. After 22:48 UT, it
moves back westward. The elongation motion pattern is similar to
that of the HXR kernel S2, which moves eastward along the PIL from
2 Mm to 16 Mm with an average speed about 55 km s$^{-1}$ from
22:42 to 22:46 UT, and then moves back along the PIL with a mean
speed about 35 km s$^{-1}$. Analysis of the UV ribbon also
suggests a consistent perpendicular expansion motion away from the
PIL at the average speed of 9 km s$^{-1}$. This is not seen in the
HXR kernel S2 due to the very weak emission and therefore there is
large uncertainty in determining the centroid of the kernel, as
can be seen from the large fluctuation in the source position.

From the analysis above, we conclude that, first, the HXR sources
and UV ribbon fronts show similar apparent motion patterns and
motion speeds, further confirming the conclusion that hard X-ray
emission and instantaneous UV brightening most likely come from
the same location on the ribbon. On the other hand, spatially
resolved UV emission exhibits a long smooth decay after the
passage of HXR kernels. Therefore, our analysis suggests that, for
this two-ribbon flare, the combined effect of elongation motion at
the measured speed and the long decay time scale explains
formation of the extended UV ribbons, whereas HXR emission
appearing only as kernels may experience a much faster decay. The
decay of UV emission will be further investigated in the next
section. Second, it is seen that the parallel motion of the ribbon
front or kernel dominates during the rise of HXR flux, whereas the
perpendicular motion starts later and dominates during the peak of
the HXR light curve, and continues into the decay of the HXR light
curve. Such evolution pattern has been reported in a few flares
\citep{Moore01, qiu09, Qiu10}.

\subsection{Decay of UV emission}
\begin{figure}
\epsscale{.90} \plotone{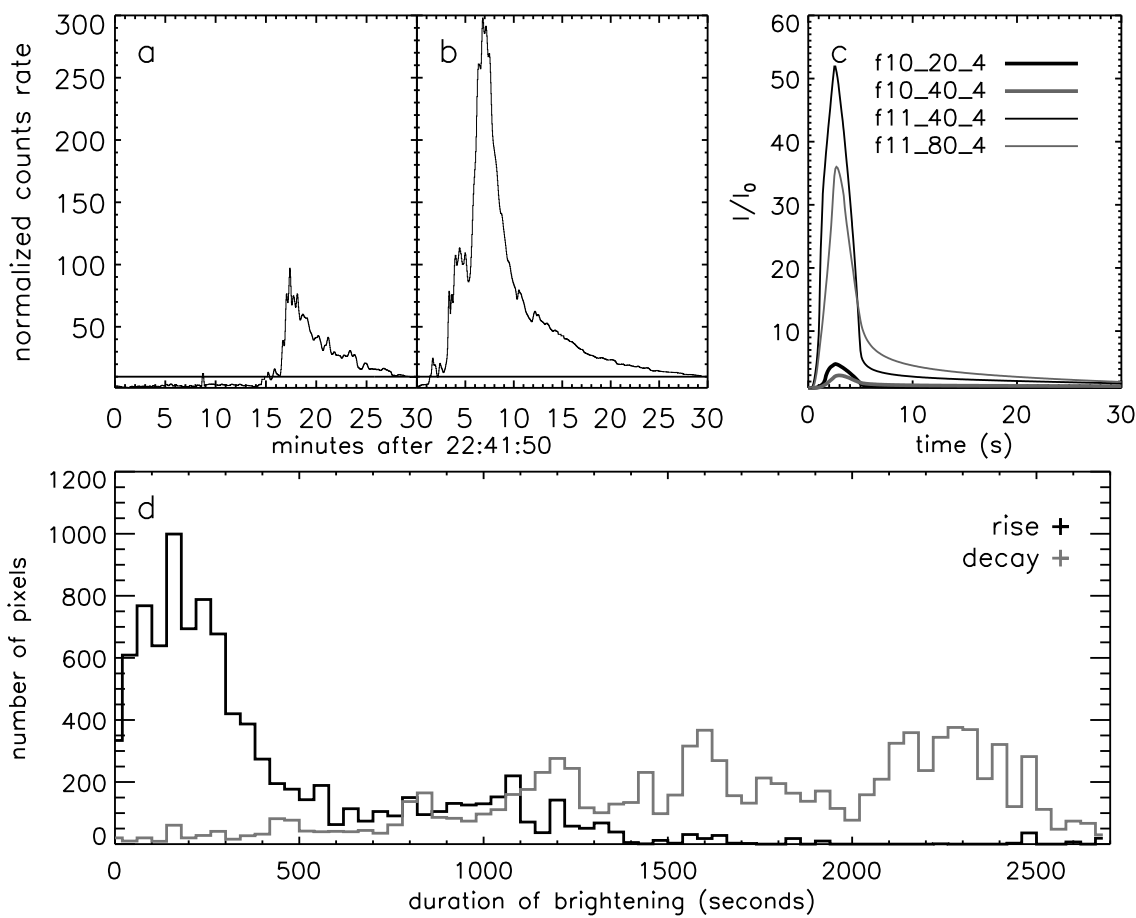} \caption{Top left: corrected UV
counts rate light curves in individual pixels. Top right: computed
UV counts rate light curves (see text) with varying electron
parameters. $f10\_20\_4$ indicates the electron beam of intensity
$f = 10^{10}$ ergs s $^{-1}$ cm$^{-2}$, electron lower energy
cutoff $\epsilon = $ 20 keV and electron spectral index $\delta =
4$. Bottom: histograms of rise and decay times of UV emissions in
8000 flaring pixels. }\label{cooling}
\end{figure}

We further investigate the long decay seen in spatially resolved UV light curves in Figure~\ref{box}.
Such a long decay may be caused by very gradual cooling of the atmosphere that contributes
to emission in the TRACE 1600\AA\ band, or by continuous heating into the decay of the flare.
Figure~\ref{cooling}a shows the light curves of a few typical pixels in 1600\AA, all demonstrating
a rapid rise and a long decay. This was previously reported by \citet{Qiu10}
in the Bastille-day flare. However, for the Bastille-day flare, the TRACE observing cadence is low,
of 30 - 40 seconds. The flare studied in this paper was observed with the highest cadence of 2~s,
so that we can determine more precisely the rise and decay times of UV counts in each individual
pixel.

In this paper, different from \citet{Qiu10}, we define the rise time as the time it takes
for the counts rate in a pixel to rise from 10 times the quiescent median to the maximum,
and the decay time is defined by the duration it takes for a pixel to decay from the maximum
to 10 times the quiescent median. Empirically, the counts rate of 10 times the quiescent median is
used as the criterion for flaring pixels \citep{Qiu10}. The histograms of the so defined rise and decay times are shown
in Figure~\ref{cooling}d. It is seen that in individual flaring pixels, the rise time ranges
from a few seconds to 4 minutes, whereas the typical ``cooling" time is over 20 minutes, about an order of magnitude
longer than the rise time. With this extended decay time and an average elongation speed of
40 km s$^{-1}$,  the ribbon spreads to a length of over 40 Mm. This is comparable with the observed ribbon length.
So, similar to \citet{Qiu10}, we conclude that the UV extended ribbon is a combined effect of
energy release in flux tubes formed by reconnection sequentially along the PIL and the long decay of UV emission.

The elongated yet apparently smooth decay in UV emission may
indicate a very long cooling process. To understand this, we
employ the dynamic radiative transfer model to compute the time
evolution of the lower atmosphere heated by non-thermal electrons
with a power-law spectrum.

We perform radiative hydrodynamics modelling using the code RADYN
\citep{carlsson92,carlsson95,carlsson97,carlsson02} with
application to solar flares as described in detail in
\citet{abbett98} and \citet{abbett99}. Atoms important to the
chromospheric energy balance are treated in non-LTE. We model
hydrogen and singly ionized calcium with six-level atoms including
the five lowest energy levels plus a continuum level. Singly
ionized magnesium is described with a four-level model atom
including the three lowest energy levels plus a continuum level.
For helium we collapse terms to collective levels and include the
$1s^2$, $2s$ and $2p$ terms in the singlet system and the $2s$ and
$2p$ terms in the triplet system of neutral helium, and the $1s$,
$2s$ and $2p$ terms of singly ionized helium. In addition we
include doubly ionized helium. We include in detail all
transitions between these levels.
 Complete frequency redistribution is assumed for
all the lines except for the Lyman transitions, in which partial
frequency redistribution is mimicked by truncating the profiles at
10 Doppler widths \citep{milkey73}. Other atomic species treated
in LTE, are included in the calculations as background continua
using the Uppsala opacity package of \citet{gus73}. Optically thin
radiative cooling due to bremsstrahlung and coronal metals is
included in the equation of internal energy conservation via an
additional cooling term.

The coupled, nonlocal, and nonlinear equations of radiative
hydrodynamics, together with the charge conservation equation, are
solved implicitly on an adaptive grid via Newton-Raphson
iteration. The required linearization of the transfer equation
follows the prescription of \citet{scharmer81}  and
\citet{scharmer85}. The adaptive mesh is that of \citet{dorfi87}
and is of critical importance when modeling the dynamics of the
lower atmosphere during flares. Features in the chromosphere, such
as strong shocks and compression waves, develop quickly and
require dense spatial distributions of grid points in order to be
properly resolved. Even with the adaptive grid, it is necessary to
use a total of 191 grid points (along with five angle points and
up to 100 frequency points) to properly resolve important
atmospheric features. Advected quantities are treated using the
second-order upwind technique of \citet{leer77}. Further details
on the basic aspects of the numerical code can be found in
\citet{carlsson97, carlsson92}  and \citet{abbett98}.

Then we compute the UV continuum radiation and convolves it with
the TRACE 1600\AA\ band response function. Figure~\ref{cooling}c
shows the computed count rates light curve as would be observed by
TRACE 1600\AA\ . These light curves are computed using different
sets of electron parameters. The electron beam used in the
simulation is defined by the peak non-thermal beam flux $f$, the
electron power-law spectral index $\delta$, and the low cutoff
energy $\epsilon_{cut}$. The simulation is conducted with $\delta
= 3, 4, 5$, $f = 10^{9}, 10^{10}, 10^{11}$ ergs cm$^{-2}$ s$^{-1}$
(denoted as $f_{9}$, $f_{10}$, and $f_{11}$, respectively), and
$\epsilon_{cut} = 20, 40, 80$ keV, respectively. The time profile
of the beam is a sinusoidal function with the rise and decay of
2.5 seconds each. The simulation shows that the result is least
sensitive to changing $\delta$. The figure shows the computed
counts rate light curve, normalized to the pre-flare quiescent
intensity, for a few cases. It is seen that at $f = 10^{11}$ ergs
cm$^{-1}$ s$^{-1}$ and $\epsilon_{cut} = 40$ keV, the maximum
enhancement is one and half order of magnitude above the pre-flare
emission. Furthermore, UV continuum emission rises nearly
simultaneously with the beam with a delay less than 1 s, and the
decay is almost as fast as the rise. With a strong beam of $\delta
= 4$, and $f = 10^{11}$ ergs cm$^{-2}$ s$^{-1}$, the decay
experiences a second gradual phase which lasts for a few minutes.
Therefore, the continuum cooling is substantially faster than
observed.

However, the computation does not take into account the C IV line
which contributes significantly to the TRACE 1600\AA\ band
\citep{Handy98}. The C~IV line is a transition region line with a
characteristic temperature of 100,000 K degree, and is therefore
formed at transition region or low corona, which is heated during
the flare. The long decay in the UV pixel is most likely dominated
by ``cooling" in the C IV line or the transition region, which
cools down slowly due to maintained coronal pressure
\citep{Fisher85, Hawley94, Griffiths98} minutes after the initial
energy deposit. {\bf The spectroscopic information is not available
for this flare to study the contribution by the C IV line
emission. However, spectroscopic observations of stellar
flares have shown that C IV line emission dominates
the decay phase of the flare emission, whereas the UV continuum
radiation has significantly decreased \citep{Hawley92}.}

\subsection{Connectivity and Energetics}
The results above suggest that magnetic reconnection and subsequent energy release take place
in individual flux tubes other than in a 2.5-dimensional arcade structure as in the standard flare model.
The brightest UV pixels at given times are most likely where instantaneous energy deposit takes place,
and where hard X-ray emission is also produced. Like many other observations, in this flare, it is also seen that
only two hard X-ray kernels are located, which are reasonably considered to be conjugate foot-points
of newly formed flare loops. In the same spirit, we assume that the brightest UV kernels in the positive and
negative magnetic fields are two conjugate foot-points of newly formed flare loops. The analysis reveals
the apparent motion pattern of the kernels, which may reflect the changing orientation of the newly formed
postflare loop. In this section, we study the orientation of the postflare loop,
assumed to be determined by the inclination between the line connecting the conjugate footpoints
and the local PIL.

\begin{figure}
\epsscale{.80} \plotone{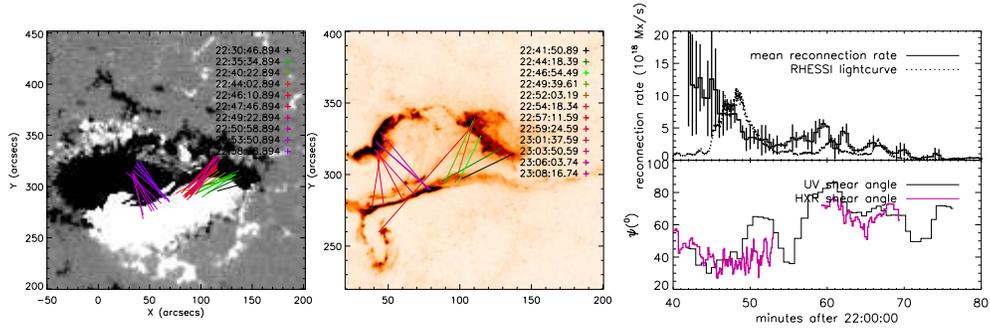} \caption{Left: Conjugate HXR
footpoints overlayed on the MDI magnetogram. Middle: conjugate UV
footpoints overlayed on a TRACE image. Right (bottom): the shear
angle of newly formed post-flare loops inferred from HXR (purple)
and UV (black) sources, respectively. Right (up):  the HXR flux in
50-100 keV (dotted line), and the reconnection rate (solid line).
Vertical bars on the reconnection rate plot indicates
uncertainties in the measurement.}\label{shear}
\end{figure}

The left panel in Figure~\ref{shear} illustrates the post-reconnection connectivity determined
from HXR kernels, and the middle panel shows the post-reconnection connectivity
determined from brightest UV emissions. The latter, with a much higher spatial resolution, would
in principle yield a more accurate measurement of the positions of the kernels. We define
the inclination angle of this projected post-reconnection loop with respect to the PIL as the
shear angle $\psi$, and plot evolution of the shear angle in the right panel in Figure~\ref{shear}.
A smaller $\psi$ indicates that the post-flare loop is more inclined toward the PIL, and a large $\psi$
indicates that the post-flare loop is more perpendicular to the PIL. It is
seen that, in general, the evolution of the shear angle determined from HXR and UV kernels is consistent.
In both plots, the shear angle continuously decreases toward the peak of HXR emission.
Around the peak of HXR emission at 22:48~UT, the shear angle is smallest (40 degrees),
indicating that the post-reconnection flare loop is most inclined toward the local PIL.
After the peak, the shear angle increases and peaks during the decay phase, consistent with the
observation that the apparent perpendicular motion proceeds into the decay phase whereas
the parallel elongation motion has ceased. After the major HXR peak, it is seen that flare
energy release takes place at a different location of the active region. The shear angle is greater,
and gradually decrease towards HXR peaks around 23:00~UT. Therefore, there is a phenomenological
relationship between flare non-thermal energetics and the orientation of the post-flare loop with
respect to the local PIL.

{\bf A plausible explanation for the anti-correlation between the hard X-ray
emission and the shear angle may involve a 3-dimentional reconnection
configuration. In this scenario, reconnection takes place between
two field lines that are not entirely anti-parallel but there is a
component of the magnetic field along the reconnection current
sheet or the polarity inversion line \citep[e.g.][]{Longcope10}. This current-aligned magnetic
field component is the so-called guide field that does not
change during reconnection. A stronger guide field would cause
the post-reconnection loop more inclined toward the polarity inversion line,
or smaller shear angle of the post-flare loop. The guide field
present in the reconnection current sheet may trap electrons
in the current sheet for longer time. If electrons are primarily
accelerated by the reconnection electric field \citep[e.g.][]{Litvinenko96},
they may be accelerated for a longer time when there is a guide field.
This scenario may explain why strong hard X-ray emission is related with
small shear angle (large guide field).}

The figure also shows the reconnection rate in terms of
reconnection flux per unit time. The reconnection rate is measured
by summing up magnetic flux in newly brightened UV pixels
\citep{Fletcher01, Qiu04, Saba06, Qiu10}. An automated procedure
has been developed to identify flaring pixels and minimize
measurement uncertainties caused by fluctuations in photometry
calibration and by non-flaring signatures \citep{Qiu10}. Displayed
is the reconnection rate averaged in positive and negative fields.
The vertical bars show the variations between the rates measured
in positive and negative fields. For this flare, there is a large
imbalance (30\%) between positive and negative reconnection rates
\citep{Fletcher01}. In this flare, since UV observations do not
cover the early phase, we miss the rise phase of reconnection
rate. The measurement shows that the reconnection rate peaks ahead
of hard X-ray emission, and then decreases monotonically. Similar
trend was reported in a few two-ribbon flares \citep{qiu09,
Qiu10}. It is noted that reconnection rate measurement takes into
account magnetic flux in all newly brightened pixels. The large
reconnection rate at the start of the flare is a result of a large
number of flaring pixels before the peak of HXR emission. This is
the stage when flare ribbon elongation dominates by kernels moving
rapidly along the PIL. The lack of correlation between the
reconnection rate and hard X-ray emission suggests that
non-thermal energy release rate per unit reconnected flux is not
uniform during the flare. During the phase of parallel motion, it
appears that reconnection is not energetically favorable, similar
to the conclusion by \citet{Qiu10}. The comparison suggests that,
not only how much flux is reconnected, but also the pattern of
reconnection, inferred from the kernel motion pattern and
post-reconnection connectivity, governs the efficiency of
non-thermal energy release (and therefore electron acceleration).

\section{Conclusions}
We analyze hard X-ray and UV observations of the two-ribbon flare on 2005 January 15
to examine the temporal and spatial relationship between emissions in these
two wavelengths and to investigate how energy release is related to magnetic
reconnection. The hard X-ray emission concentrates in two kernels in opposite magnetic fields
whereas UV emission appears as two ribbons. The UV ribbon fronts and HXR kernels both exhibit apparent
impulsive parallel motion and steady and slow perpendicular motion with respect to the magnetic PIL.
It is evident that, along both ribbons, UV emission is impulsively enhanced
when and where HXR kernel sweeps through. After the passage of the HXR kernel, UV emission decays gradually
on timescales of over 20 minutes, apparently undergoing an elongated cooling.
The analysis suggests that significant UV emission is primarily produced by non-thermal beam deposit
which also produces hard X-ray emission. {\bf When such relationship can be established,
it is possible to measure the area of non-thermal electron precipitation more precisely
by combining high-resolution imaging UV observations with hard X-ray observations. This will provide
an observational constraint on non-thermal beam parameters, such as the beam flux.}

Furthermore, the observations of apparent motion of
instantaneous energy release location and the long cooling time of UV emission provide an explanation for the
apparent morphological discrepancy in the two wavelengths, that UV emission appears as extended ribbons
whereas hard X-ray emission appears as compact kernels. These results support the scenario that
reconnection and subsequent energy release takes place in individual flux tubes nearly
sequentially along the magnetic polarity inversion line. The observations of this two-ribbon
flare therefore presents the picture of 3D reconnection different from the 2.5D arcade configuration
in the standard flare model.

\begin{figure}
\epsscale{1.20} \plotone{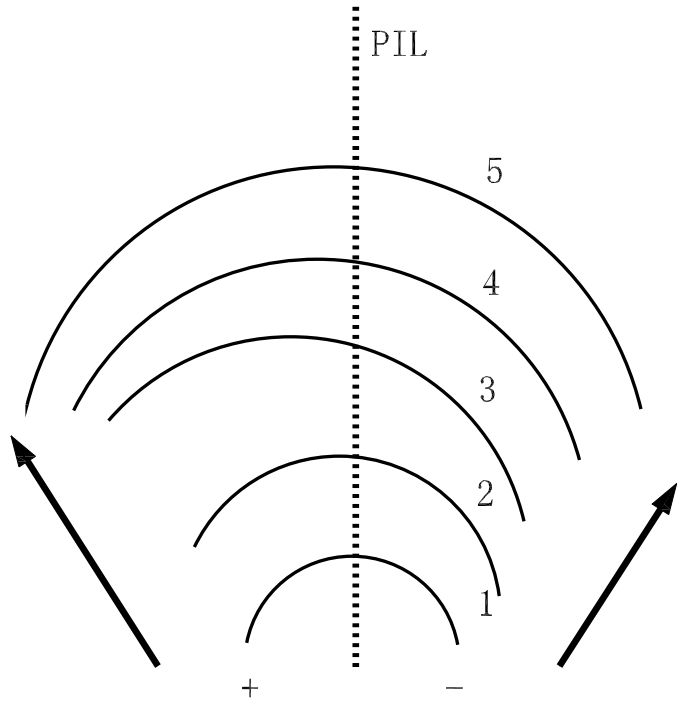} \caption{A sketch of the possible
post-flare loop arcade configuration, giving the top view of the
sequentially formed post-flare loops. Low-lying loops (indicated
in the figure as loops 1, 2, and 3) are formed earlier along the
arcade, with the shear with the polarity inversion line (dashed
line) decreasing with time. High-lying loops (loops 4 and 5) are
formed later with increasing shear. The two arrows indicate the
observed ribbon motion direction. The configuration explains the
observed dominant parallel motion of the flare ribbon during the
rise phase and the dominant perpendicular motion
afterwards.}\label{cartoon}
\end{figure}

It is also observed that the apparent parallel motion of the hard
X-ray kernels and UV fronts dominates during the rise phase of
hard X-ray emission, when reconnection rate peaks, and
perpendicular motion dominates around the peak of the emission,
when non-thermal energy flux is thought to reach maximum. With a
2.5D approximation, we also infer the inclination of the newly
formed post-flare loops with respect to the PIL. The analysis
shows the the shear angle of the postflare loop with respect to
the PIL is smallest during the hard X-ray emissions, consistent
with the motion pattern observed during the evolution of the
flare. These observations suggest that efficiency of non-thermal
energy release (in terms of non-thermal energy flux per unit
reconnected flux) is related to the manner of reconnection as
inferred from the inclination of newly formed post-flare loops. It
is likely that the amount of magnetic energy release per unit
reconnection flux is different depending on the shear change from
the pre-flare to post-flare configuration. It is, however, also
plausible that the inclination of post-flare loops indicates
existence of a guide field component in the reconnection
\citep{Qiu10}. The guide field is the magnetic field component in
the current sheet that is parallel with the reconnection electric
field and does not participate in reconnection. Existence of the
guide field modifies reconnection physics, and is particularly
important in particle energization \citep[e.g.;][]{Litvinenko96},
which may be reflected in the observed phenomenological
relationship between reconnection (apparent motion) pattern and
non-thermal energetics. {\bf In this scenario, evolution of the shear
angle is determined by pre-reconnection magnetic field configuration, such
that low-lying field lines that reconnect earlier are becoming less sheared, and
then the overlying field lines that reconnect later are more sheared, as shown
in the cartoon in Figure~\ref{cartoon}.}

Finally, we have also conducted the dynamic radiative transfer simulation to investigate evolution of UV continuum emission
produced by non-thermal beams with varying beam parameters. It is found that the UV continuum emission
time profile follows closely the heating function, which decays rapidly as soon as the heating terminates.
The elongated cooling observed in the TRACE 1600\AA\ band emission most likely reflects the transition region
emission in the C IV line. In the future work, the C IV line contribution will be computed to provide a better
guide in diagnosing radiative signatures observed in TRACE 1600 \AA\ band. This will aide the analysis
of energetics in flux tubes formed and energized by reconnection.

We thank Drs. R. W. Nightingale and T. Tarbell for help with TRACE
calibration. We acknowledge TRACE, SoHO, and RHESSI missions for
providing quality observations. This work is supported by NASA
grant NNX08AE44G and NSF grant ATM-0748428. Part of the work was
conducted during the NSF REU program at Montana State University
and supported by Dr. D. E. McKenzie.

{}

\clearpage

\begin{figure}
\epsscale{.80} \plotone{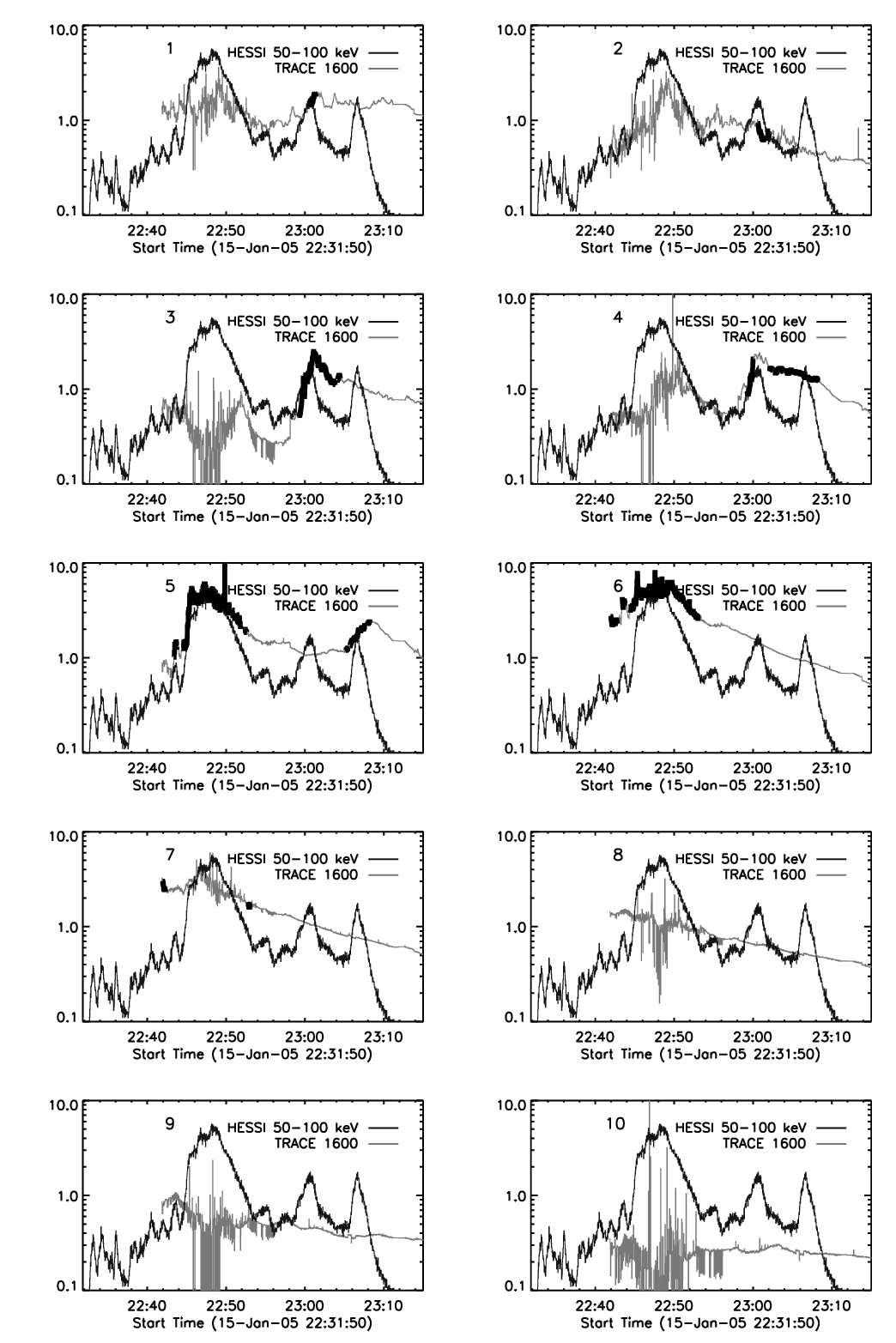} \caption{The corrected TRACE UV
counts rate light curves (grey) in small boxes as indicated in
Figure 2 (P-ribbon), compared with HXR 50-100 keV count flux light
curve (dark) arbitrarily normalized. The thick dark color on a
TRACE light curve shows the timing when the HXR source passes
through a certain box.}
\end{figure}

\clearpage

\end{document}